\documentclass[pra,a4paper,aps,twocolumn,superscriptaddress,nofootinbib,10pt]{revtex4-2}

\usepackage{graphicx}
\usepackage{dcolumn}
\usepackage{bm,bbm}
\usepackage{xcolor}
\usepackage{tcolorbox}
\usepackage{algorithm}
\usepackage{algpseudocode}
\usepackage{amsmath}
\usepackage{bbold}
\usepackage{braket}
\usepackage{nameref}
\usepackage[toc,page]{appendix}

\usepackage[colorlinks,breaklinks,linkcolor={blue},citecolor={blue},urlcolor={blue}]{hyperref}
\usepackage{enumerate}
\usepackage{physics}

\newcommand{\qo}[1]{``#1''}   

\begin{document}

\title{Generation of the Complete Bell Basis via Hong-Ou-Mandel Interference}

\author{Xiaoqin Gao}
\thanks{Current affiliation: National Laboratory of Solid State Microstructures and School of Physics, Collaborative Innovation Center of Advanced Microstructures, Jiangsu Physical Science Research Center, Nanjing University, Nanjing, Jiangsu 210093, China}
\affiliation{Nexus for Quantum Technologies, University of Ottawa, K1N 5N6, Ottawa, ON, Canada}

\author{Dilip Paneru}
\affiliation{Nexus for Quantum Technologies, University of Ottawa, K1N 5N6, Ottawa, ON, Canada}

\author{Francesco Di Colandrea}\email{francesco.dicolandrea@unina.it}
\affiliation{Nexus for Quantum Technologies, University of Ottawa, K1N 5N6, Ottawa, ON, Canada}
\affiliation{Dipartimento di Fisica, Universit\`{a} degli Studi di Napoli Federico II, Complesso Universitario di Monte Sant'Angelo, Via Cintia, 80126 Napoli, Italy}

\author{Yingwen Zhang}\email{yzhang6@uottawa.ca}
\affiliation{Nexus for Quantum Technologies, University of Ottawa, K1N 5N6, Ottawa, ON, Canada}
\affiliation{National Research Council of Canada, 100 Sussex Drive, K1A 0R6, Ottawa, ON, Canada}

\author{Ebrahim Karimi}
\affiliation{Nexus for Quantum Technologies, University of Ottawa, K1N 5N6, Ottawa, ON, Canada}
\affiliation{National Research Council of Canada, 100 Sussex Drive, K1A 0R6, Ottawa, ON, Canada}
\affiliation{Institute for Quantum Studies, Chapman University, Orange, California 92866, USA}

\date{\today}

\begin{abstract}
Optical vector modes (VMs), characterized by spatially varying polarization distributions, have become essential tools across microscopy, metrology, optical trapping, nanophotonics, and optical communications. The Hong-Ou-Mandel (HOM) effect, a fundamental two-photon interference phenomenon in quantum optics, offers significant potential to extend the applications of VMs beyond the classical regime. Here, we demonstrate the simultaneous generation of all four Bell states by exploiting the HOM interference of VMs. The resulting Bell states exhibit spatially tailored distributions that are determined by the input modes. These results represent a significant step in manipulating HOM interference within structured photons, offering promising avenues for high-dimensional quantum information processing and in particular high-dimensional quantum communication, quantum sensing, and advanced photonic technologies reliant on tailored quantum states of light.
\end{abstract}
\maketitle

Hong-Ou-Mandel (HOM) interference is a two-particle quantum interference effect and plays a central role in many quantum technologies that require interaction between two particles~\cite{PhysRevLett.59.2044,bouchard2020two}. When two indistinguishable photons enter a 50:50 beamsplitter from different input ports, the photons destructively interfere and experience \qo{bunching}, in which they exit through the same output port of the beamsplitter, resulting in the well-known dip in coincidence counts. If the two input photons are entangled, then a HOM dip in the coincidences will be observed if the two photons' entangled state is symmetric, and a HOM peak will be observed if the entangled state is antisymmetric. HOM interference is widely used in photonic quantum experiments, such as the implementation of quantum logic gates~\cite{RevModPhys.79.797}, quantum imaging~\cite{PhysRevLett.91.083601,Ndagano2022}, quantum communication~\cite{Hu2023}, and optimal quantum cloning~\cite{PhysRevLett.92.047902, 2009, bouchard2017high}. The effect is also used to characterize photon-pair sources~\cite{PhysRevA.98.053811}, optimize quantum teleportation~\cite{Bennett1993}, and enable Boson sampling experiments~\cite{zhong2020quantum, PhysRevLett.127.180502}.

Vector modes (VMs), also known as vector vortex modes, possess a distinctive spatially varying polarization structure that differentiates them from homogeneously polarized light beams~\cite{zhan2009cylindrical}. This characteristic has allowed its application in various fields, including nonlinear optics~\cite{PhysRevLett.90.013903}, high-resolution imaging~\cite{Chen:07}, and optical trapping~\cite{Kozawa:10,Moradi:19}. VMs also offer enhanced capabilities, such as sharper focus spots~\cite{PhysRevLett.91.233901,wang2008creation}, and improved performance in quantum key distribution~\cite{souza2008quantum,PhysRevLett.113.060503}. Two primary methods for generating VMs are the intra-cavity method, which incorporates mode-selection optical elements within a laser resonator~\cite{sakai2007optical, miyai2006lasers,Iwahashi:11}, and the extra-cavity method, which employs specialized optical components, such as $q$-plates~\cite{PhysRevLett.96.163905,Cardano:12, Cardano:13} or spatial light modulators within Sagnac interferometers, to convert Gaussian beams into VMs~\cite{Maurer_2007}. The entanglement of VMs has also been explored, where polarization-entangled photon pairs are converted into VMs, creating a complex, spatially varying entanglement structure~\cite{PhysRevA.82.022115,fickler2013real,DAmbrosio2016,Graffitti2020,PhysRevLett.132.063802}. 

Recent studies on HOM interference with VMs have revealed spatially distributed HOM dips and peaks~\cite{schiano2024engineering}. However, these studies are restricted to two spatial dimensions, focusing on the spatially dependent polarization structure of photons from only one of the interferometer's output ports, captured with an event camera, while photons in the other port are detected with a bucket detector. By correlating time-stamped events between the camera and the bucket detector, a two-dimensional map of the final-state structure was constructed. Nevertheless, the complete structure of the VMs after HOM interference is inherently four-dimensional as it depends on the spatial coordinates of photons from both output ports.

Here, we demonstrate a full spatial characterization of HOM interference with VMs, revealing the complete four-dimensional state structure. Remarkably, unlike previous studies suggesting that only the antisymmetric Bell state $\psi^-$ can emerge from HOM interference when post-selecting on coincidence events between the two output ports, we find that interference between VMs can simultaneously yield all four polarization Bell states, depending on the relative transverse spatial locations of detected photons in the two output ports. We believe that this finding has significant implications for quantum technologies that leverage HOM interference.

\begin{figure}[t]
\includegraphics [width= 0.48\textwidth]{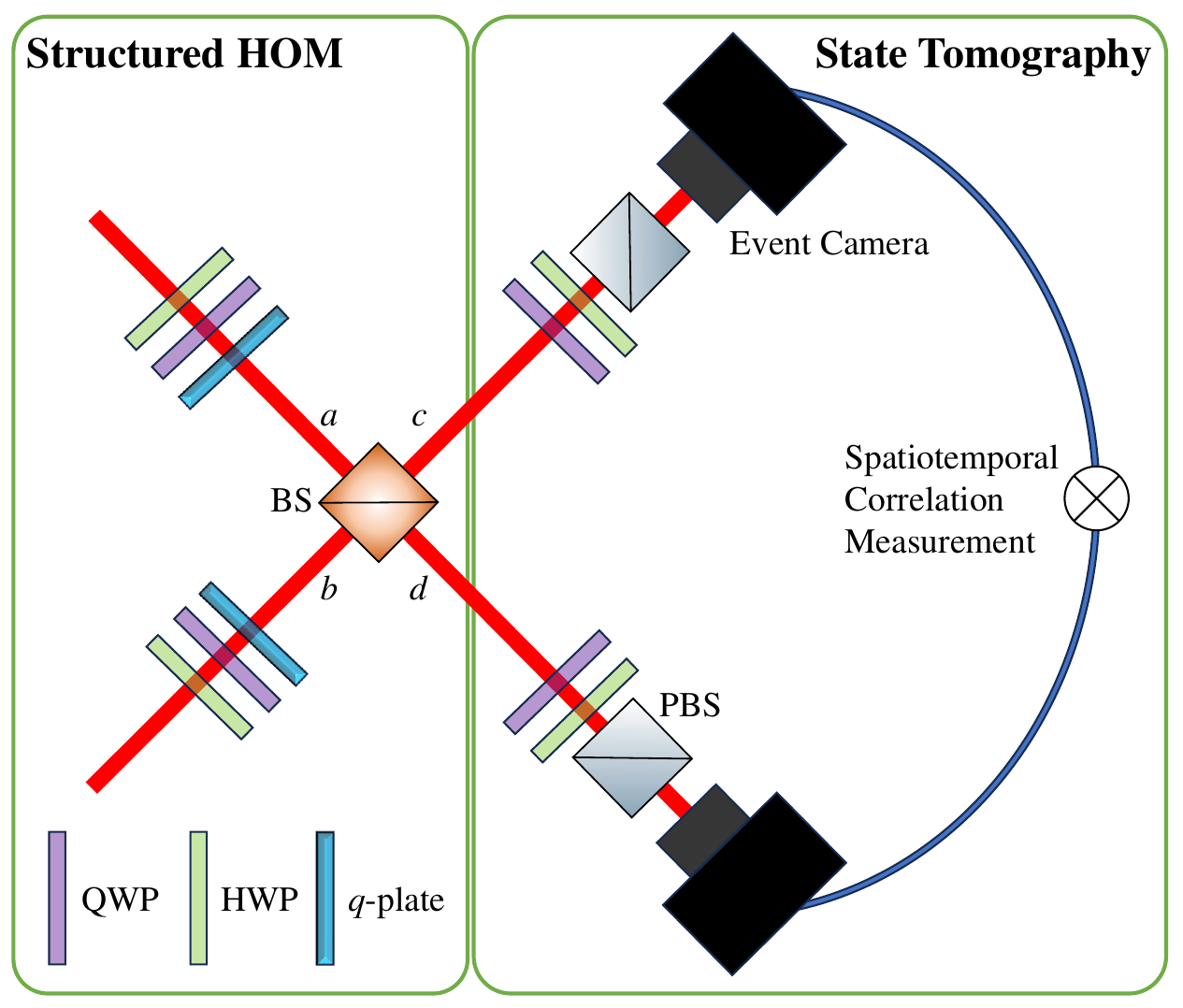}
\centering
\caption{
{\textbf{Space-resolved HOM interference of VMs.} Single Gaussian modes are sent to two $q$-plates to create VMs. These overlap at a 50:50 BS, where HOM interference occurs. Space-resolved polarization state tomography is performed by recording the outputs of two projection stages, consisting of a QWP, a HWP, and a polarizing beamsplitter (PBS), onto two event cameras (in the actual experiment, two different regions of a single camera were used instead). By extracting temporal-spatial correlated events between all pixels of the two cameras, a spatially varying density matrix can be retrieved, from which the distribution of the Bell states is obtained.}
}
\label{fig:concept}
\end{figure}
A simplified experimental scheme for realizing HOM interference of VMs is shown in Fig.~\ref{fig:concept}; the detailed experimental setup can be found in the Supplementary Material. Photon pairs are generated from Spontaneous Parametric Down-Conversion (SPDC) in a type-II ${5\text{-mm-long}}$ ppKTP crystal and coupled to single-mode fibers (not shown in the figure) to select only the Gaussian mode. Two sets of quarter-wave plates (QWPs) and half-wave plates (HWPs) are used to prepare the input polarization states: 
\begin{equation}
\ket{\psi_0}=a^{\dagger}_{\pi_1}b^{\dagger}_{\pi_2}\ket{0},
\end{equation}
where ${a^{\dagger}_{\pi_1}}$ (${b^{\dagger}_{\pi_2}}$) is the creation operator for a photon in input port $a$ ($b$) with polarization ${\ket{\pi_1}}$ (${\ket{\pi_2}}$). The two photons are then sent to two $q$-plates. In the circular polarization basis, where ${\ket{L}}$ and ${\ket{R}}$ are left-handed and right-handed circular polarization states, the action of a $q$-plate with topological charge $q$ and birefringence $\delta$ on an input Gaussian mode is~\cite{karimi2009light,slussarenko2011tunable}
\begin{equation}
\begin{split}
    \hat{U}_q^\delta\cdot\left[ \begin{array}{ll}
        \ket{L}\\
        \ket{R}\\
    \end{array} \right] &= F_{0}(r)\cos\left(\frac{\delta}{2}\right)\left[ \begin{array}{ll}
        \ket{L}\\
        \ket{R}\\
    \end{array} \right] \\&
  + i \sin\left(\frac{\delta}{2}\right)F_{q}(r)\left[ \begin{array}{ll}
       e^{-i2q\theta}\ket{R}\\
      e^{i2q\theta}\ket{L}\\
    \end{array} \right],
    \end{split}
    \label{eqn:qplateeq}
\end{equation}
where ${r}$ and ${\theta}$ are the radial and azimuthal coordinate in the transverse plane, respectively, and $F_{0}(r)$ and $F_{q}(r)$ are Hypergeometric-Gauss functions~\cite{Karimi:09}, which can be approximated as the Laguerre-Gauss mode with orbital angular momentum index ${\ell=2q}$ and radial index equal to zero: ${F_{q}(r)e^{\pm i2q\theta}\equiv\text{LG}_{\pm 2q,0}(r,\theta)}$.

The birefringence parameter ${\delta}$ is uniform across the plate and can be electrically controlled~\cite{Piccirillo2010}. In this experiment, we set ${\delta=\pi}$, which corresponds to a fully tuned $q$-plate. Interference between VMs takes place within a 50:50 beamsplitter (BS), whose action, neglecting a global phase, reads
\begin{equation}
\begin{split}
\ket{\psi}_a \xrightarrow[]{\mathrm{BS}}\frac{1}{\sqrt{2}}\left(\ket{\psi}_c+\ket{\psi}_d \right);\\
\ket{\psi}_b \xrightarrow[]{\mathrm{BS}}\frac{1}{\sqrt{2}}\left(\ket{\psi}_c-\ket{\psi}_d \right),
\end{split}
\label{eqn:BSplit}
\end{equation}
where $c$ and $d$ are the output ports, and the transformation applies to all degrees of freedom. Assuming two horizontally polarized input photons, ${\ket{\psi_0}=a^\dagger_Hb^\dagger_H\ket{0}}$, with ${\ket{H}=\left(\ket{L}+\ket{R}\right)/\sqrt{2}}$, and after post-selecting only on coincidence events between output ports $c$ and $d$, the resulting state after the BS can be written as (see the Supplementary Material for the detailed derivation)
\begin{equation}
\begin{split}
\ket{\psi_f} = &\mathcal{N}\left(s_{\phi^+}\ket{\phi^+} + s_{\phi^-}\ket{\phi^-}\right.\\
&\left. +s_{\psi^+}\ket{\psi^+} + s_{\psi^-}\ket{\psi^-}\right) ,    
\end{split}
\label{eqn:finalstate}
\end{equation}
where ${\mathcal{N}}$ is a normalization constant, $\ket{\phi^\pm}=\left(\ket{HH}\pm\ket{VV}\right)/\sqrt{2}$; $\ket{\psi^\pm}=\left(\ket{HV}\pm\ket{VH}\right)/\sqrt{2}$ are the four polarization Bell states, and $s_{\phi^\pm}$; $s_{\psi^\pm}$ are given by
\begin{eqnarray}
    s_{\phi^+} &=&F_{q_a}(r_d)F_{q_b}(r_c)\cos(2q_a\theta_d-2q_b\theta_c)\cr
    &-&F_{q_a}(r_c)F_{q_b}(r_d)\cos(2q_a\theta_c-2q_b\theta_d);\\
    s_{\phi^-}&=&F_{q_a}(r_d)F_{q_b}(r_c)\cos(2q_a\theta_d+2q_b\theta_c)\cr
    &-&F_{q_a}(r_c)F_{q_b}(r_d)\cos(2q_a\theta_c+2q_b\theta_d);\\
    s_{\psi^+}&=&F_{q_a}(r_d)F_{q_b}(r_c)\sin(2q_a\theta_d+2q_b\theta_c)\cr
    &-&F_{q_a}(r_c)F_{q_b}(r_d)\sin(2q_a\theta_c+2q_b\theta_d);\\
    s_{\psi^-}&=&F_{q_a}(r_d)F_{q_b}(r_c)\sin(2q_a\theta_d-2q_b\theta_c)\cr
    &+&F_{q_a}(r_c)F_{q_b}(r_d)\sin(2q_a\theta_c-2q_b\theta_d),
\end{eqnarray}
with ${q_a}$ (${q_b}$) the topological charge of the $q$-plate in input port ${a}$ (${b}$). 

To experimentally characterize this state, the output photons are sent to a polarization tomography apparatus consisting of a QWP, a HWP, and a polarizing beamsplitter (PBS) in each output arm, and are then directed to two different regions of an event camera (TPX3CAM)~\cite{timepix1,timepix2}. The size of the beam spot on the camera covers approximately $30\times30$ pixels. Photon pairs are identified using time correlation measurements between the camera pixels. The space-resolved outcomes of 16 polarimetric measurements are processed with a maximum-likelihood method to retrieve a density matrix ${\rho(r_c,\theta_c,r_d,\theta_d)}$ at each pixel~\cite{PhysRevA.64.052312}, from which the probability distributions of the four Bell states are extracted:
\begin{equation}
\begin{split}
P_{\phi^{\pm}}(r_c,\theta_c,r_d,\theta_d)&=\mel{\phi^\pm}{\rho(r_c,\theta_c,r_d,\theta_d)}{\phi^\pm};  \\
P_{\psi^{\pm}}(r_c,\theta_c,r_d,\theta_d)&=\mel{\psi^\pm}{\rho(r_c,\theta_c,r_d,\theta_d)}{\psi^\pm}. 
\end{split}
\end{equation}

\begin{figure*}[t]
\includegraphics [width= 1\textwidth]{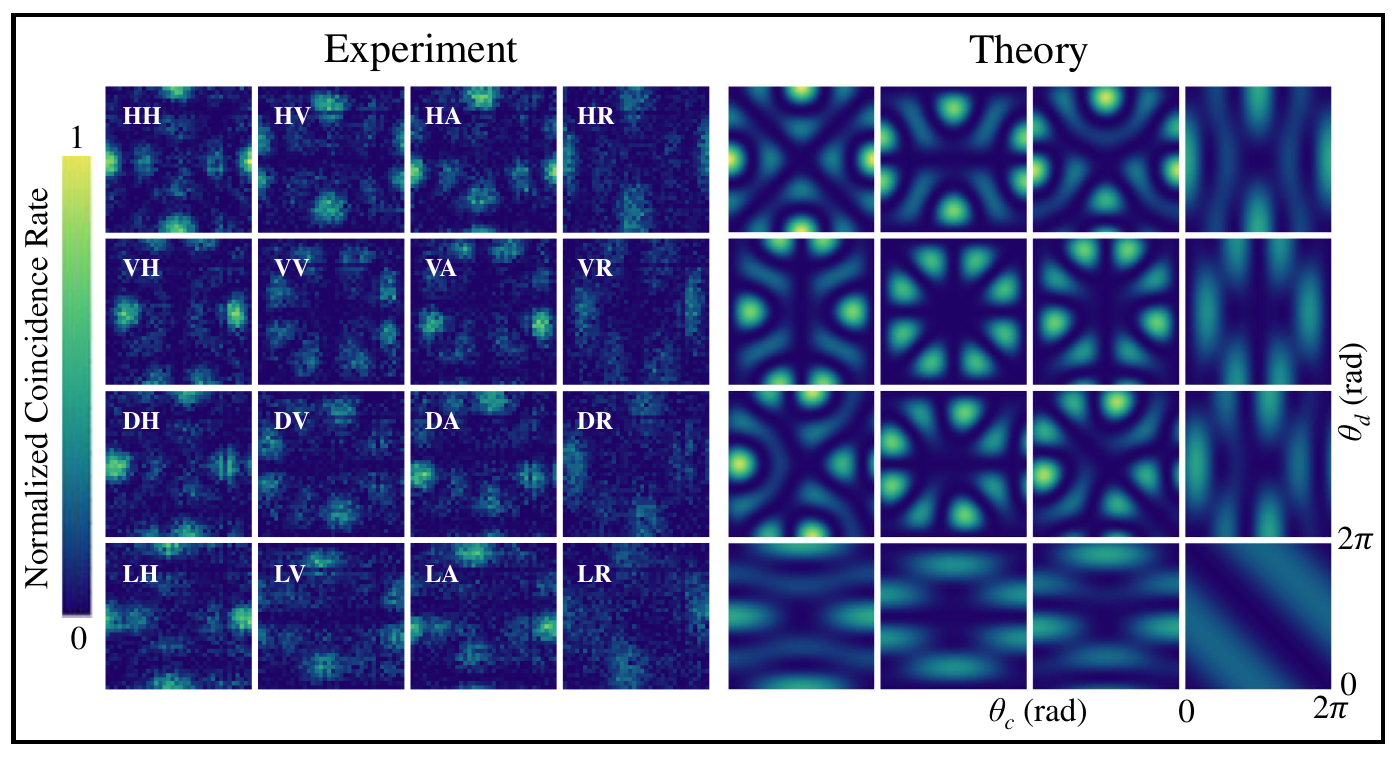}
\centering
\caption{{\textbf{Space-resolved polarimetry of HOM interference of VMs.} Experimental (left) and theoretical (right) azimuthal correlations of the two-photon state for the 16 projections required for the complete state tomography. The topological charges of the two $q$-plates are ${q_a = 1}$ and ${q_b = 1/2}$. The initial state is $\hat{a}_{H}^\dagger\hat{b}_{H}^\dagger\ket{0}$.}}
\label{prob1&1/2}
\end{figure*}
\begin{figure*}[t]
\includegraphics [width=1\linewidth]{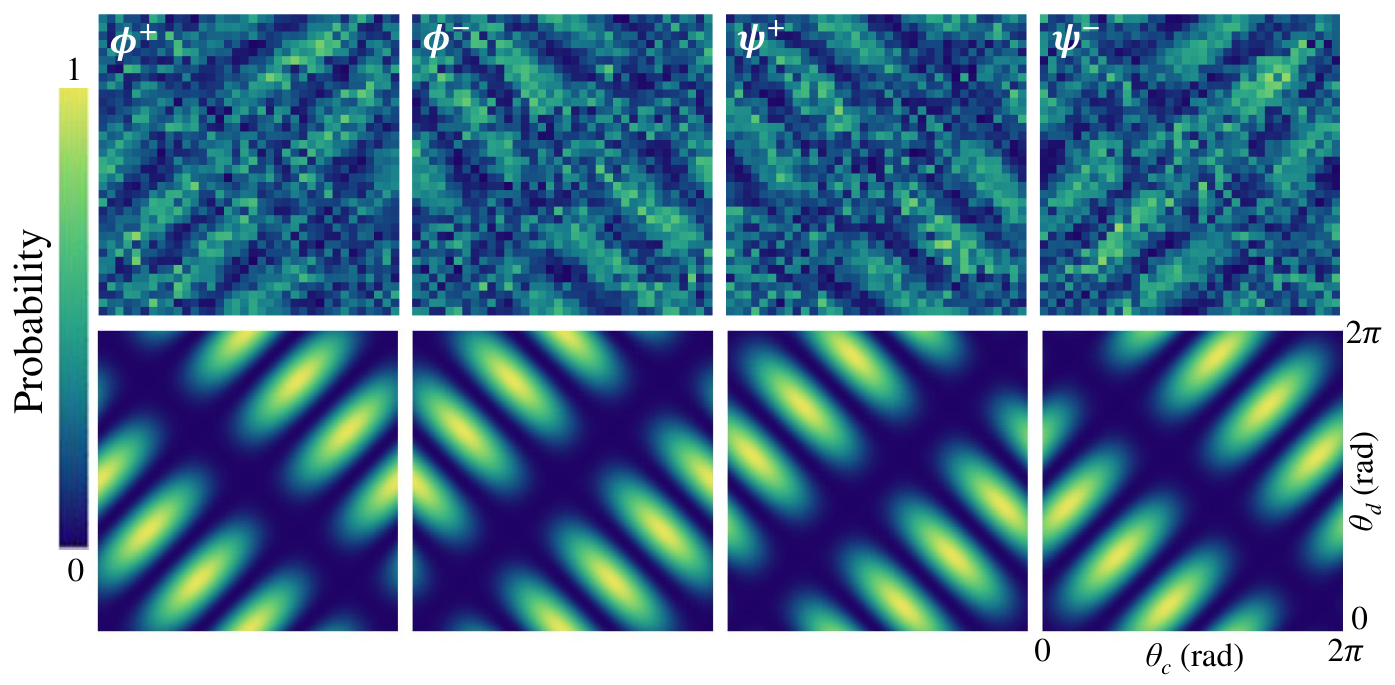}
\centering
\caption{\textbf{Probability distributions of the four Bell states.} Experimental (top) and theoretical (bottom) spatial patterns of the four Bell states obtained via HOM interference of VMs for the case study ${q_a=1}$, ${q_b=1/2}$, and initial state  ${\hat{a}_{H}^\dagger\hat{b}_{H}^\dagger\ket{0}}$. 
}
\label{bell}
\end{figure*}
Figure~\ref{prob1&1/2} shows the complete set of 16 polarization measurements from the two output ports with the two $q$-plates having topological charges ${q_a = 1}$ and ${q_b= 1/2}$. Due to the cylindrical symmetry of the Laguerre-Gauss modes in which the final state showed minimal variations with respect to the radial coordinates, therefore, we only plotted the azimuthal correlations for each polarization projection, obtained as the marginal distributions ${P(\theta_c,\theta_d)=\sum_{r_c,r_d}P(r_c,\theta_c,r_d,\theta_d)}$. The comparison with theoretical predictions (grayscale), calculated from Eq.~\eqref{eqn:finalstate}, shows good agreement in all polarization projections. 

From the reconstructed marginal density matrix, ${\rho(\theta_c,\theta_d)}$, we extracted the corresponding space-dependent Bell state decomposition, as shown in Fig.~\ref{bell}. Contrary to the typical HOM effect, where only the $\psi^-$ state is expected to be observed when conditioned on coincidence events between the two output ports, the output states of VM HOM feature contributions from all Bell states, depending on the specific location in the coordinate space. The locations where each Bell state can be uniquely found require the following conditions be met:
\begin{equation}
\begin{split}
    \text{only}~P_{\phi^+}>0 ~\text{iff} &\quad (q_a+q_b)(\theta_c-\theta_d) = \frac{(2n+1)\pi }{2}\\
    &\quad \text{and} \quad (q_a-q_b)(\theta_c-\theta_d) = m\pi;\\
    \text{only}~P_{\phi^-}>0 ~\text{iff} &\quad (q_a+q_b)(\theta_c+\theta_d) = \frac{(2n+1)\pi }{2}\\
    &\quad \text{and} \quad (q_a-q_b)(\theta_c+\theta_d) = m\pi;\\
    \text{only}~P_{\psi^+}>0 ~\text{iff} &\quad (q_a+q_b)(\theta_c+\theta_d) = n\pi\\
    &\quad \text{and} \quad (q_a-q_b)(\theta_c+\theta_d) = m\pi;\\
    \text{only}~P_{\psi^-}>0 ~\text{iff} &\quad (q_a+q_b)(\theta_c-\theta_d) = n\pi\\
    &\quad \text{and} \quad (q_a-q_b)(\theta_c-\theta_d) = m\pi,
\end{split}
\label{conditions1}
\end{equation}
with $n,m$ being an integer. From this, we also obtain the following conditions on $q_a$ and $q_b$:
\begin{equation}
\begin{split}
    \text{only}~P_{\phi^+}>0 ~\text{iff} &\quad q_a = \frac{(2n+2m+1)}{(2n-2m+1)}q_b ;\\
    \text{only}~P_{\phi^-}>0 ~\text{iff} &\quad q_a = \frac{(2n+2m+1)}{(2n-2m+1)}q_b ;\\
    \text{only}~P_{\psi^+}>0 ~\text{iff} &\quad q_a = \frac{(n+m)}{(n-m)}q_b;\\
    \text{only}~P_{\psi^-}>0 ~\text{iff} &\quad q_a = \frac{(n+m)}{(n-m)}q_b;\\
    \text{for all cases} &\quad \abs{q_a}\neq \abs{q_b}.
\end{split}
\label{conditions2}
\end{equation}
In the case where ${q_a = 1}$ and ${q_b= 1/2}$, the conditions are satisfied for only $P_{\psi^+}>0$  if and only if $\theta_c = -\theta_d$ and for only $P_{\psi^-}>0$  where $\theta_c = \theta_d$; however, the conditions cannot be satisfied for only $P_{\phi^+}>0$ or only $P_{\phi^-}>0$. The smallest $q_a$ and $q_b$ values where all four Bell states can be uniquely found are when $q_a = \pm 3/2$ and $q_b = \mp 1/2$.
For more details on the derivation of the conditions in Eq.~\eqref{conditions1}, see the Supplementary Material. Polarization measurements and Bell state decomposition for different $q$-plates settings and input polarization states can also be found in the Supplementary Material.


In conclusion, we exploited VMs to imprint a complex spatial structure on the outcome of HOM interference, resulting in a distinct spatial distribution composed of all four Bell states. We also showed where within this spatial distribution can each Bell state be uniquely found. We provided theoretical and experimental evidence of this effect in the case of VMs generated by $q$-plates, which naturally induce an azimuthal structure on the final distribution. However, this functionality can be easily generalized to arbitrary spatial patterns~\cite{Larocque_2016}, allowing the distribution of Bell states of the output to be tailored to specific demands. For instance, liquid-crystal polarization gratings ($g$-plates)~\cite{DErrico:20} could be used to map each Bell state onto a two-dimensional Cartesian grid. The results of this work could have promising uses in applications such as quantum communications and quantum sensing. In the near future, it would be interesting to extend similar concepts to a higher number of input photons, where the controlled generation of multi-particle maximally entangled states remains extremely challenging~\cite{PhysRevLett.121.250505,Cogan2023,osullivan2024}.  



\section*{Acknowledgments}
The authors would like to thank Ryan Hogan for valuable discussions. This work was supported by the Canada Research Chairs (CRC) and NRC-uOttawa Joint Centre for
Extreme Quantum Photonics (JCEP) via the Quantum Sensors Challenge Program at the National Research Council of Canada. Xiaoqin Gao would like to thank the support of the Natural Science Foundation of Jiangsu Province (No.~BK20233001). Francesco Di Colandrea acknowledges support from the PNRR MUR project PE0000023-NQSTI.

\bibliographystyle{apsrev4-1fixed_with_article_titles_full_names_new}
\bibliography{refs}

\clearpage
\onecolumngrid
\renewcommand{\figurename}{\textbf{Figure}}
\setcounter{figure}{0} \renewcommand{\thefigure}{\textbf{S{\arabic{figure}}}}
\setcounter{table}{0} \renewcommand{\thetable}{S\arabic{table}}
\setcounter{section}{0} \renewcommand{\thesection}{S\arabic{section}}
\setcounter{equation}{0} \renewcommand{\theequation}{S\arabic{equation}}
\onecolumngrid

\begin{center}
{\Large Supplementary Material for: \\ Generation of the Complete Bell Basis via Hong-Ou-Mandel Interference}
\end{center}
\vspace{1 EM}

\section{Experimental setup}

\begin{figure*}[h]
\includegraphics [width= 1\textwidth]{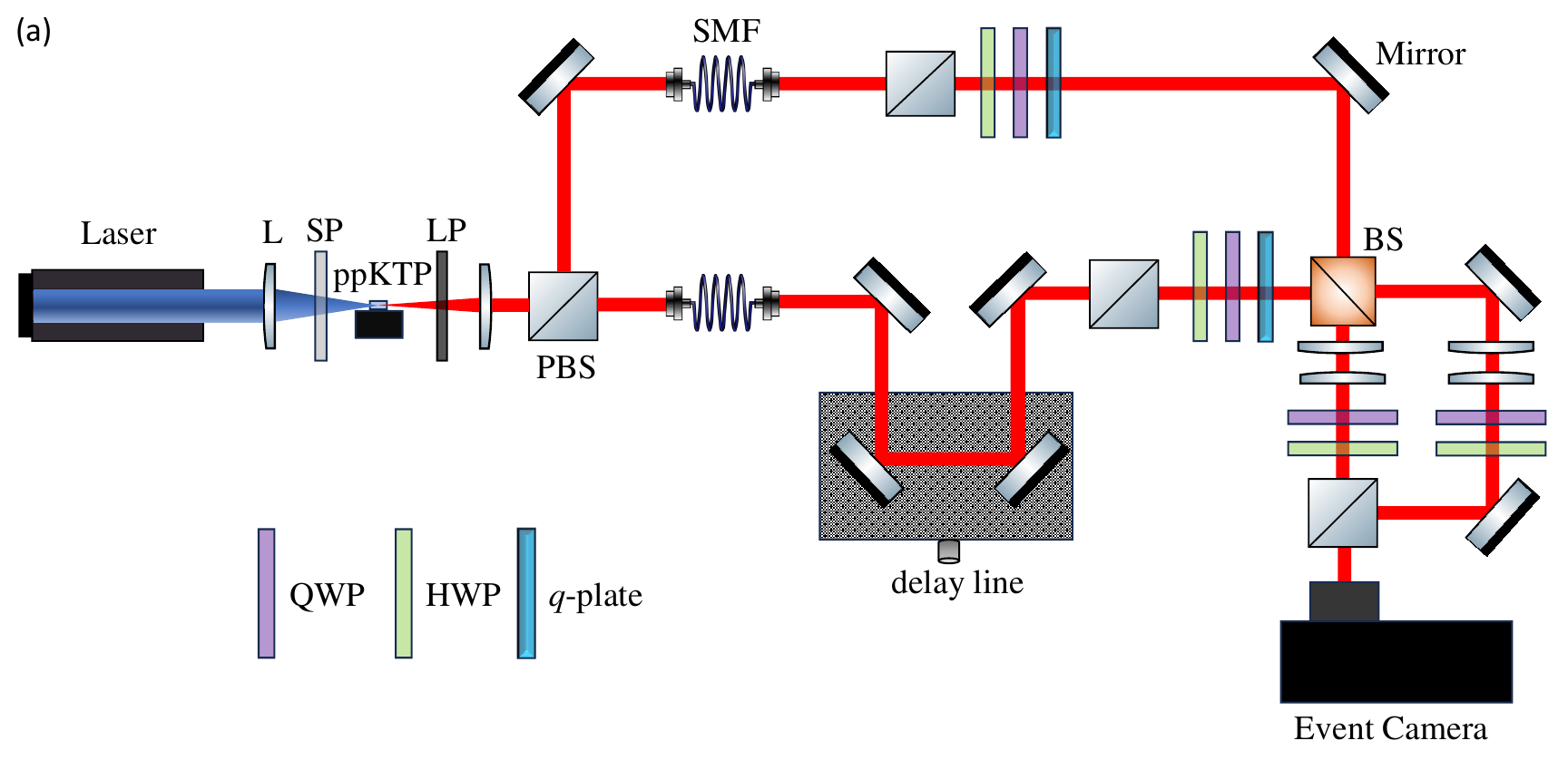}
\includegraphics [width= 1\textwidth]{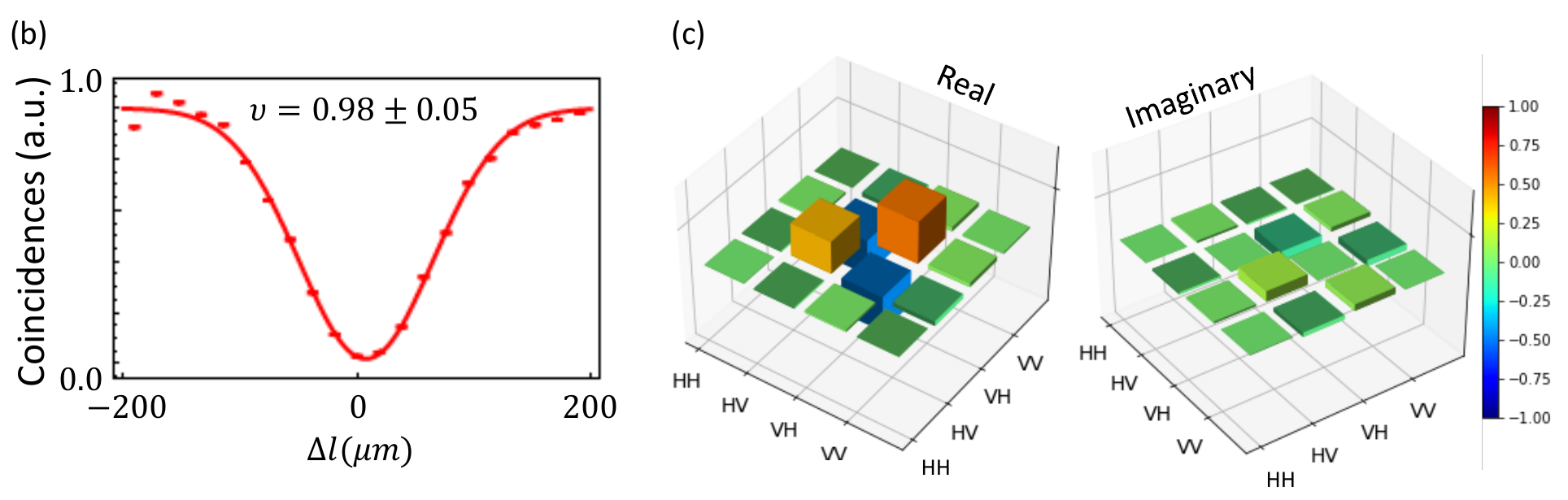}
\centering
\caption{{\bf Generating Bell states via structured HOM interference.} {(a)~}A 405\, nm \eqref{eqn:qplateeq} continuous-wave laser pumps a Type-II 5-mm-long ppKTP crystal generating photon pairs with orthogonal polarization. These are first coupled into SMFs for spatial-mode filtering. Two stages HWP-QWP set the input polarization. $q$-plates are used to transform the spatially uniform polarization into VMs. Polarization state tomography is performed using a combination QWP-HWP-PBS, after which each photon is imaged onto different regions of the event camera, where spatiotemporal information of the output photons is recorded. (b)~When $q$-plates are turned off, a HOM dip with a visibility ${v=0.98\pm0.05}$ is obtained when the initial state is $\hat{a}_{H}^\dagger\hat{b}_{H}^\dagger\ket{0}$ 
(c)~Polarization density matrix (real and imaginary part) reconstructed for the antisymmetric Bell state $\ket{\psi^-}$, which is obtained when the initial state is $\hat{a}_{V}^\dagger\hat{b}_{H}^\dagger\ket{0}$. L - Lens, SP - Shortpass filter, LP - Longpass filter. 
}
\label{fig:setup}
\end{figure*}

The experimental setup for the generation and characterization of Bell states from Hong-Ou-Mandel (HOM) interference of vector modes (VMs) is shown in Fig.~\ref{fig:setup}(a). A 405~nm continuous-wave laser pumps a Type-II 5-mm-long ppKTP crystal generating photon pairs with orthogonal polarization through Spontaneous Parametric Down-Conversion (SPDC). The photons are separated through a polarizing beamsplitter (PBS) and coupled to single-mode fibers (SMFs) to filter out only the Gaussian mode contribution. Two sets of half-wave plates (HWPs) and quarter-wave plates (QWPs) allow adjusting their input polarization onto two $q$-plates which generate the desired VMs, depending on the imprinted topological charges. The photons are then sent to a 50:50 beamsplitter (BS) to undergo HOM interference. A delay line is added to one arm of the HOM interferometer to control the optical path difference ${\Delta l}$, to ensure temporal overlap of the photons at the BS. After the photons pass the BS, a ${4f}$ system is used to image the output modes onto two separate areas of the event camera (TPX3CAM). Prior to detection, the photons pass through another set of QWP-HWP-PBS performing polarization projective measurements for state tomography. 

Temporal overlap was checked by first turning off the $q$-plates (${\delta=2\pi}$) and preparing two identical photons in the state $\hat{a}_{H}^\dagger\hat{b}_{H}^\dagger\ket{0}$, for which a HOM dip with visibility ${v=0.98\pm0.05}$ was observed (see Fig.~\ref{fig:setup}(b)). 
The visibility of the HOM dip in photon bunching is defined as $v_\text{dip}=(C-C_{min})/C$, where $C_{min}$ is the minimum in-dip coincidence count and $C$ is the out-of-dip coincidence count, where the difference in the two path lengths is larger than the coherence length of the SPDC photons.
An alternative check to the quality of HOM interference is provided in Fig.~\ref{fig:setup}(c), which shows the polarization density matrix (real and imaginary part) of the antisymmetric Bell state $\ket{\psi^-}$ 
obtained by preparing the state $\hat{a}_{V}^\dagger\hat{b}_{H}^\dagger\ket{0}$ as the input, with ${\ket{V}}$ the vertical polarization state, and post-selecting on coincidence events between the two output ports. Here, a state purity of $\sim 99\%$ was observed. 

\clearpage
\section{Derivation of the final state}

Here, the derivation is for assuming an input state of ${\ket{\psi_0}=a^\dagger_Hb^\dagger_H\ket{0}}$; the derivation for other input states can be derived in a similar way. 

The $q$-plates in each arm transform the ${\ket{\psi_0}=a^\dagger_Hb^\dagger_H\ket{0}}$ state as follows (see Eq.~\eqref{eqn:qplateeq}):

\begin{equation}
U^{\delta_a=\pi}_{q_a}U^{\delta_b=\pi}_{q_b}\ket{H}_a\ket{H}_b=F_{q_a}(r_a)F_{q_b}(r_b)\left(\cos{2q_a\theta_a}\ket{H}_a+\sin{2q_a\theta_a}\ket{V}_a\right)\left(\cos{2q_b\theta_b}\ket{H}_b+\sin{2q_b\theta_b}\ket{V}_b\right),
\label{eqn:transformed}
\end{equation}
where ${q_a}$ and ${q_b}$ are the topological charges of the $\pi$-tuned $q$-plates in port $a$ and $b$, respectively, and ${F_q(r)}$ is the radial part of the Laguerre-Gauss mode carrying ${\ell=2q}$ units of orbital angular momentum.

The 50:50 BS operations can be written as (see Eq.~\eqref{eqn:BSplit})
\begin{equation}
\begin{split}
\ket{\psi}_a \xrightarrow[]{\mathrm{BS}}\frac{1}{\sqrt{2}}\left(\ket{\psi}_c+\ket{\psi}_d \right);\\
\ket{\psi}_b \xrightarrow[]{\mathrm{BS}}\frac{1}{\sqrt{2}}\left(\ket{\psi}_c-\ket{\psi}_d \right),
\end{split}
\end{equation}
and applying it to the transformed state of Eq.~\eqref{eqn:transformed} we obtain
\begin{equation}
\begin{split}
\ket{\psi}=\mathcal{M}& \big[F_{q_a}(r_c)\left(\cos{2q_a\theta_c}\ket{H}_c+\sin{2q_a\theta_c\ket{V}_c} \right)+F_{q_a}(r_d)\left(\cos{2q_a\theta_d}\ket{H}_d+\sin{2q_a\theta_d\ket{V}_d} \right) \big]\\
&
\big[F_{q_b}(r_c)\left(\cos{2q_b\theta_c}\ket{H}_c+\sin{2q_b\theta_c\ket{V}_c} \right)-F_{q_b}(r_d)\left(\cos{2q_b\theta_d}\ket{H}_d+\sin{2q_b\theta_d\ket{V}_d} \right) \big],
\end{split}
\end{equation}
where $\mathcal{M}$ is a normalization constant. By selecting coincidence events, i.e., the contributions that have the two photons exiting from two different ports, we obtain
\begin{equation}
\begin{split}
\ket{\psi_f}=\,&\mathcal{N}  \big(-F_{q_a}(r_c)F_{q_b}(r_d)\cos{2q_a\theta_c}\cos{2q_b\theta_d}+F_{q_a}(r_d)F_{q_b}(r_c)\cos{2q_a\theta_d}\cos{2q_b\theta_c} \big)\ket{H}_c\ket{H}_d\\
&+\big(-F_{q_a}(r_c)F_{q_b}(r_d)\sin{2q_a\theta_c}\sin{2q_b\theta_d}+F_{q_a}(r_d)F_{q_b}(r_c)\sin{2q_a\theta_d}\sin{2q_b\theta_c}\big)\ket{V}_c\ket{V}_d;\\
&+\big(-F_{q_a}(r_c)F_{q_b}(r_d)\cos{2q_a\theta_c}\sin{2q_b\theta_d}+F_{q_a}(r_d)F_{q_b}(r_c)\sin{2q_a\theta_d}\cos{2q_b\theta_c}\big)\ket{H}_c\ket{V}_d;\\
&+\big(-F_{q_a}(r_c)F_{q_b}(r_d)\sin{2q_a\theta_c}\cos{2q_b\theta_d}+F_{q_a}(r_d)F_{q_b}(r_c)\cos{2q_a\theta_d}\sin{2q_b\theta_c}\big)\ket{V}_c\ket{H}_d,   
\end{split}
\end{equation}
which can be rearranged using the trigonometric identities for angle addition and subtraction into
\begin{equation}
\ket{\psi_f} = \mathcal{N}\left(s_{\phi^+}\ket{\phi^+} + s_{\phi^-}\ket{\phi^-} +s_{\psi^+}\ket{\psi^+} + s_{\psi^-}\ket{\psi^-}\right),    
\end{equation}
where $\ket{\phi^\pm}=\frac{\ket{HH}\pm\ket{VV}}{\sqrt{2}}$; $\ket{\psi^\pm}=\frac{\ket{HV}\pm\ket{VH}}{\sqrt{2}}$ are the four Bell states and $s_{\phi^\pm}$; $s_{\psi^\pm}$ are given by
\begin{equation}
\begin{split}
s_{\phi^+} = \,&F_{q_a}(r_d)F_{q_b}(r_c)\cos(2q_a\theta_d-2q_b\theta_c) -F_{q_a}(r_c)F_{q_b}(r_d)\cos(2q_a\theta_c-2q_b\theta_d);\\
s_{\phi^-} = \,&F_{q_a}(r_d)F_{q_b}(r_c)\cos(2q_a\theta_d+2q_b\theta_c) -F_{q_a}(r_c)F_{q_b}(r_d)\cos(2q_a\theta_c+2q_b\theta_d);\\
s_{\psi^+} = \,&F_{q_a}(r_d)F_{q_b}(r_c)\sin(2q_a\theta_d+2q_b\theta_c) -F_{q_a}(r_c)F_{q_b}(r_d)\sin(2q_a\theta_c+2q_b\theta_d);\\
s_{\psi^-} = \,&F_{q_a}(r_d)F_{q_b}(r_c)\sin(2q_a\theta_d-2q_b\theta_c) +F_{q_a}(r_c)F_{q_b}(r_d)\sin(2q_a\theta_c-2q_b\theta_d),
\end{split}
\label{S6}
\end{equation}
which corresponds to the main result of Eq.~\eqref{eqn:finalstate}. By following the same approach, the interference results can be derived for different input states.  

Summing out the radial dependence, Eq.~\eqref{S6} can be rewritten using the sum-to-product trigonometric identities as
\begin{equation}
\begin{split}
s_{\phi^+} = \,&2\sin\left[(q_a-q_b)(\theta_c+\theta_d)\right]\sin\left[(q_a+q_b)(\theta_c-\theta_d)\right];\\
s_{\phi^-} = \,&2\sin\left[(q_a+q_b)(\theta_c+\theta_d)\right]\sin\left[(q_a-q_b)(\theta_c-\theta_d)\right];\\
s_{\psi^+} = &-2\sin\left[(q_a-q_b)(\theta_c-\theta_d)\right]\cos\left[(q_a+q_b)(\theta_c+\theta_d)\right];\\
s_{\psi^-} = \,&2\sin\left[(q_a-q_b)(\theta_c+\theta_d)\right]\cos\left[(q_a+q_b)(\theta_c-\theta_d)\right].
\end{split}
\end{equation}
The locations where each Bell state can be uniquely found require the following conditions:
\begin{equation}
\begin{split}
   \text{only} ~s_{\phi^+}^2>0 ~\text{iff}    &\quad (q_a+q_b)(\theta_c-\theta_d) = \frac{(2n+1)\pi }{2} \quad \text{and} \quad (q_a-q_b)(\theta_c-\theta_d) = m\pi;\\
    \text{only} ~s_{\phi^-}^2>0 ~\text{iff} &\quad (q_a+q_b)(\theta_c+\theta_d) = \frac{(2n+1)\pi }{2} \quad \text{and} \quad (q_a-q_b)(\theta_c+\theta_d) = m\pi;\\
    \text{only} ~s_{\psi^+}^2>0 ~\text{iff} &\quad (q_a+q_b)(\theta_c+\theta_d) = n\pi \quad \text{and} \quad (q_a-q_b)(\theta_c+\theta_d) = m\pi;\\
    \text{only}~s_{\psi^-}^2>0 ~\text{iff} &\quad (q_a+q_b)(\theta_c-\theta_d) = n\pi \quad \text{and} \quad (q_a-q_b)(\theta_c-\theta_d) = m\pi,
\end{split}
\label{S8}
\end{equation}

with $n,m$ being an integer. From this, we also obtain the following conditions on $q_a$ and $q_b$:
\begin{equation}
\begin{split}
    \text{only} ~s_{\phi^+}^2>0 ~\text{iff}  &\quad (2n-2m+1)q_a = (2n+2m+1)q_b ;\\
    \text{only} ~s_{\phi^-}^2>0 ~\text{iff} &\quad (2n-2m+1)q_a = (2n+2m+1)q_b ;\\
    \text{only} ~s_{\psi^+}^2>0 ~\text{iff} &\quad (n-m)q_a = (n+m)q_b;\\
    \text{only} ~s_{\psi^-}^2>0 ~\text{iff} &\quad (n-m)q_a = (n+m)q_b;\\
    \text{for all cases} &\quad \abs{q_a}\neq \abs{q_b}.
\end{split}
\end{equation}

\section{Supplementary Data}
\begin{figure*}[tbph]
\includegraphics [width= 1\textwidth]{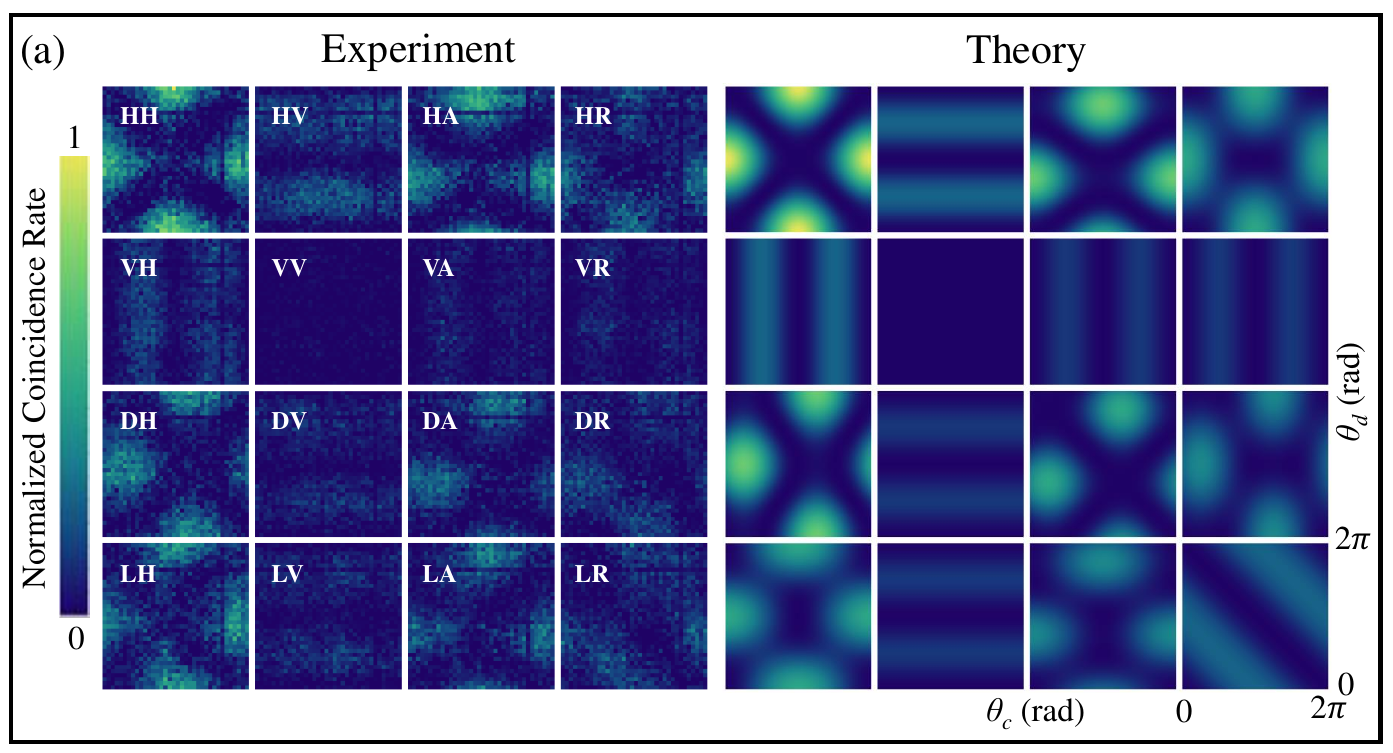}
\includegraphics [width= 1\textwidth]{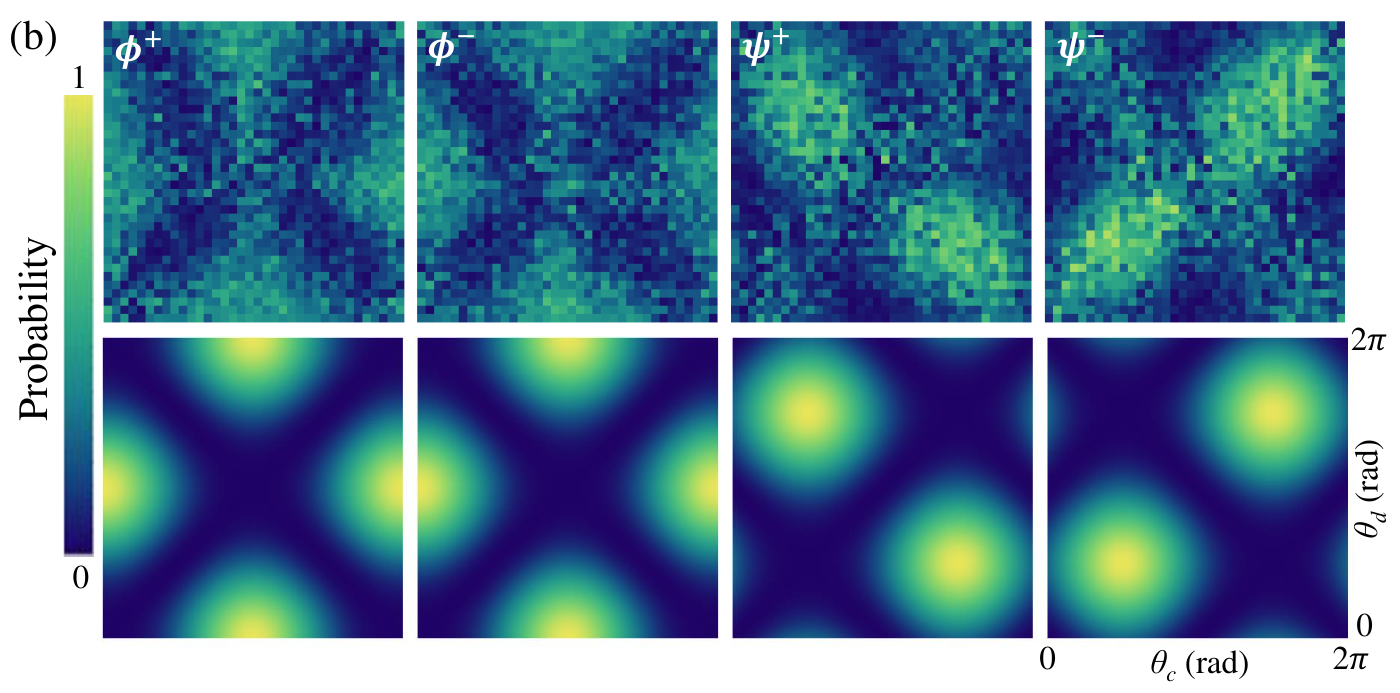}
\centering
\caption{\textbf{Generation of Bell states from HOM interference of VMs.} (a)~Experimental (left) and simulated (right) azimuthal correlations of the two-photon state for the 16 projections required for the complete state tomography. The first {$q$-plate} is turned off while the topological charge of the second $q$-plate is ${q_b = 1/2}$. The initial state is $\hat{a}_{H}^\dagger\hat{b}_{H}^\dagger\ket{0}$. (b)~Experimental (top) and simulated (bottom) spatial patterns of the four Bell states ${\lbrace \ket{\phi^+},\ket{\phi^-},\ket{\psi^+},\ket{\psi^-} \rbrace}$ obtained via HOM interference of VMs in the case study. 
}
\label{case1}
\end{figure*}

\begin{figure*}[tbph]
\includegraphics [width= 1\textwidth]{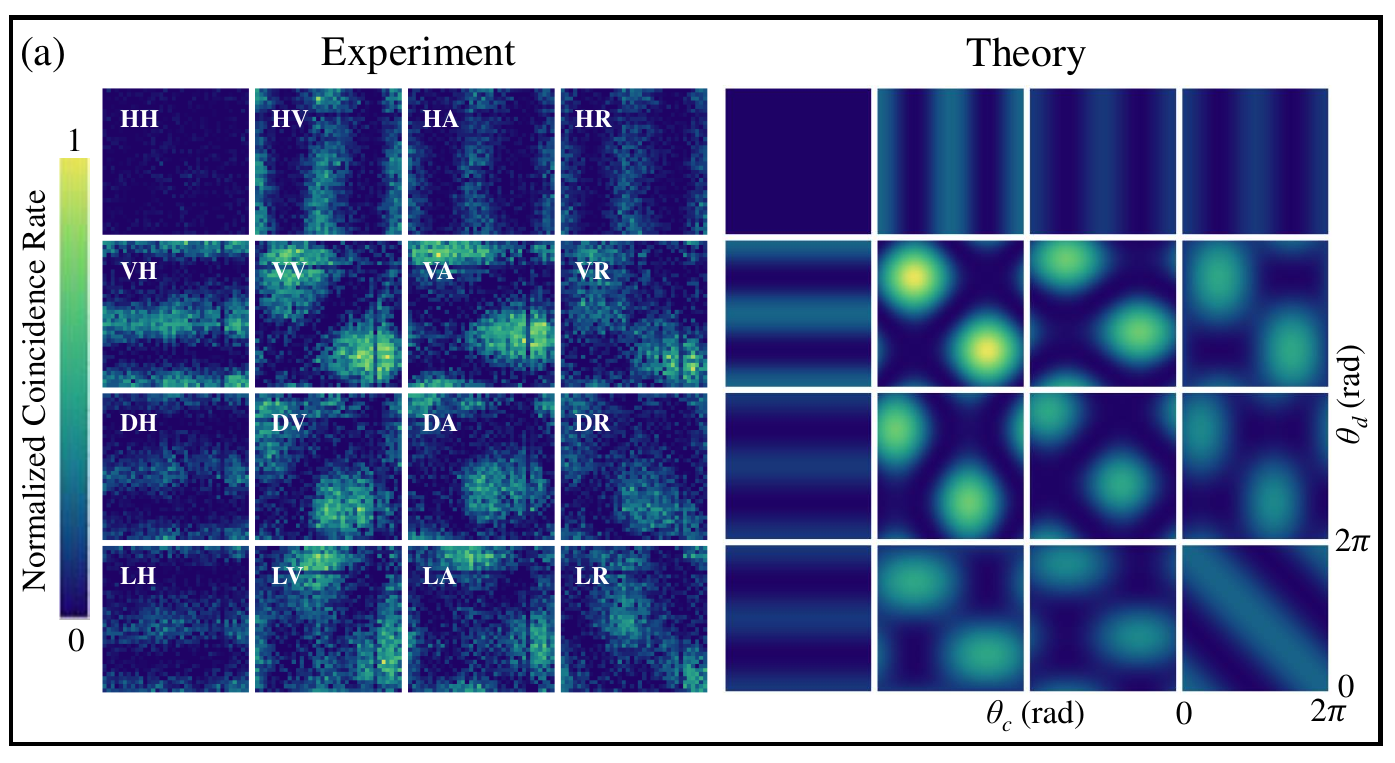}
\includegraphics [width= 1\textwidth]{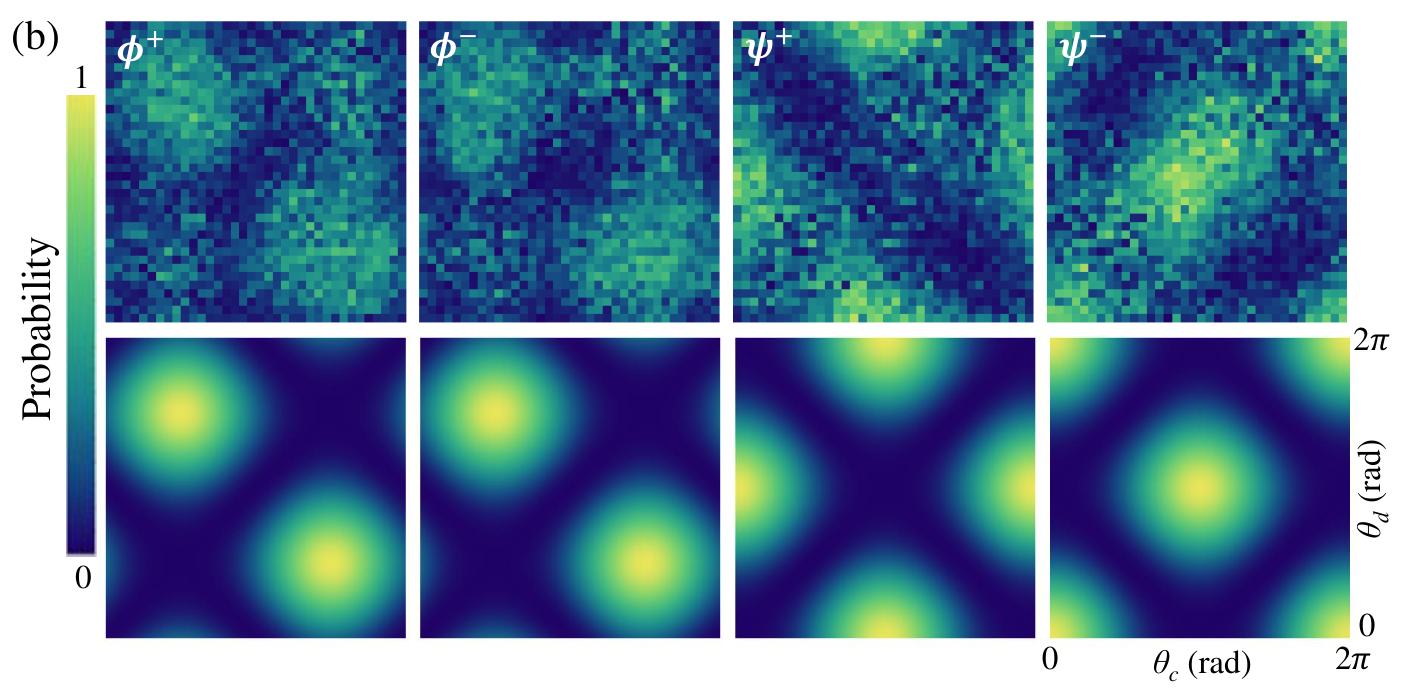}
\centering
\caption{\textbf{Generation of Bell states from HOM interference of VMs.} (a)~Experimental (left) and simulated (right) azimuthal correlations of the two-photon state for the 16 projections required for the complete state tomography. The first $q$-plate is turned off while the topological charge of the second $q$-plate is ${q_b = 1/2}$. The initial state is $\hat{a}_{V}^\dagger\hat{b}_{H}^\dagger\ket{0}$. (b)~Experimental (top) and simulated (bottom) spatial patterns of the four Bell states ${\lbrace \ket{\phi^+},\ket{\phi^-},\ket{\psi^+},\ket{\psi^-} \rbrace}$ obtained via HOM interference of VMs in the case study. 
}
\label{case2}
\end{figure*}

\begin{figure*}[tbph]
\includegraphics [width= 0.98\textwidth]{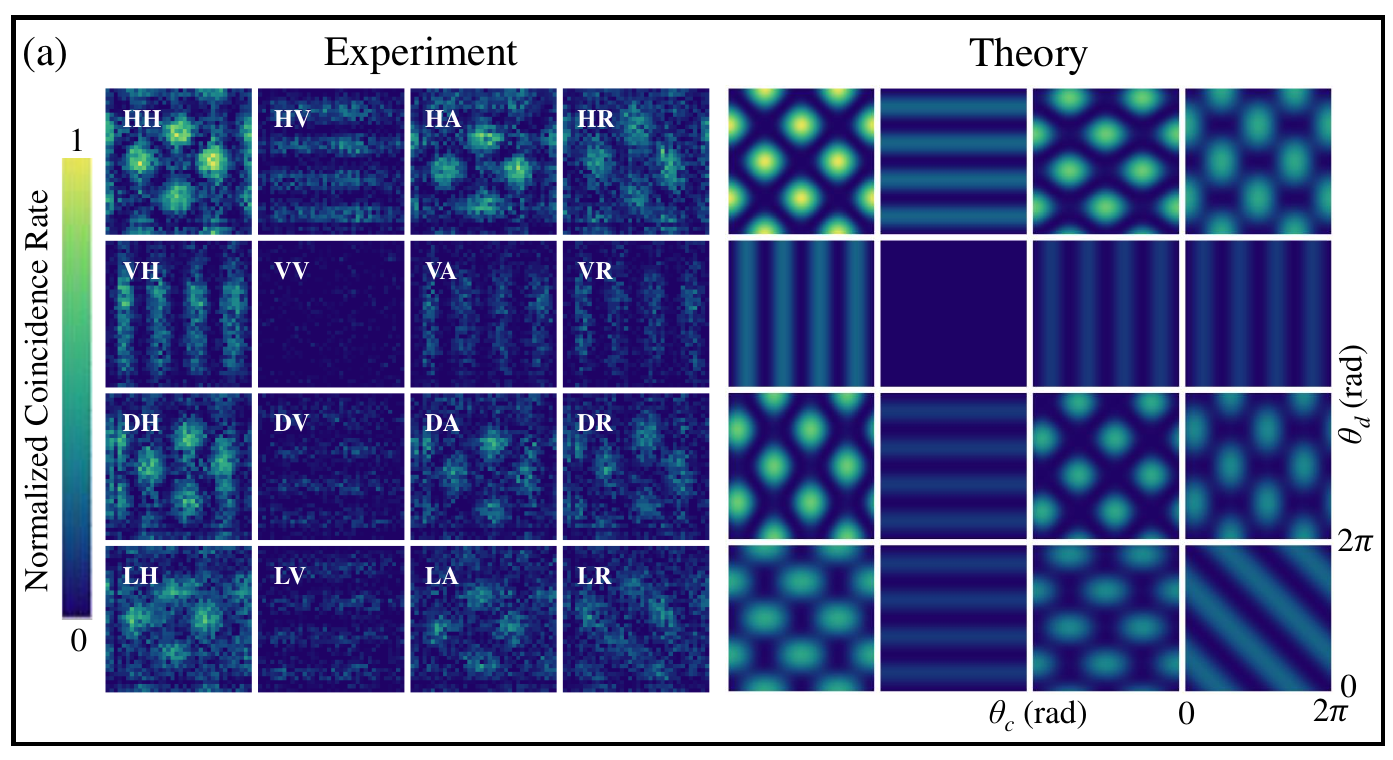}
\includegraphics [width= 1\textwidth]{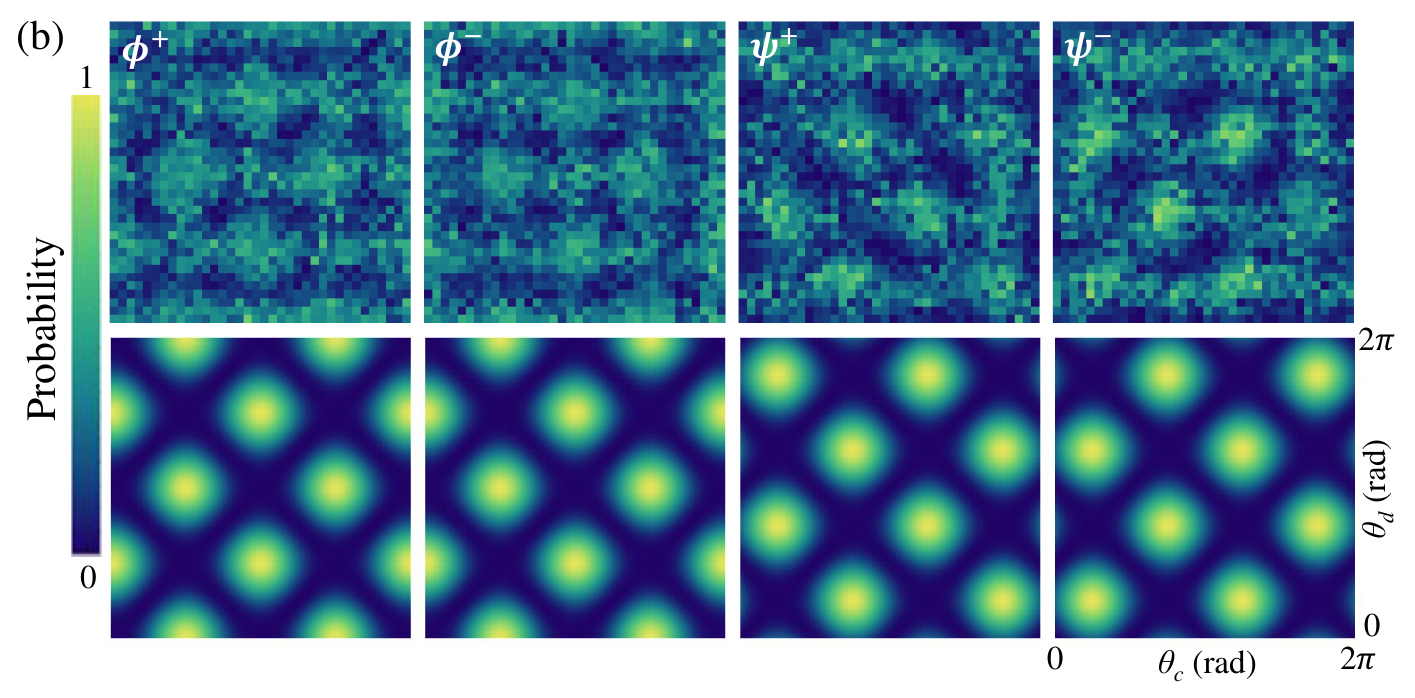}
\centering
\caption{\textbf{Generation of Bell states from HOM interference of VMs.} (a)~Experimental (left) and simulated (right) azimuthal correlations of the two-photon state for the 16 projections required for the complete state tomography. The topological charge of the first $q$-plate is ${q_a = 1}$ while the second $q$-plate is turned off. The initial state is $\hat{a}_{H}^\dagger\hat{b}_{H}^\dagger\ket{0}$. (b)~Experimental (top) and simulated (bottom) spatial patterns of the four Bell states ${\lbrace \ket{\phi^+},\ket{\phi^-},\ket{\psi^+},\ket{\psi^-} \rbrace}$ obtained via HOM interference of VMs in the case study. 
}
\label{case3}
\end{figure*}

\begin{figure*}[tbph]
\includegraphics [width= 1\textwidth]{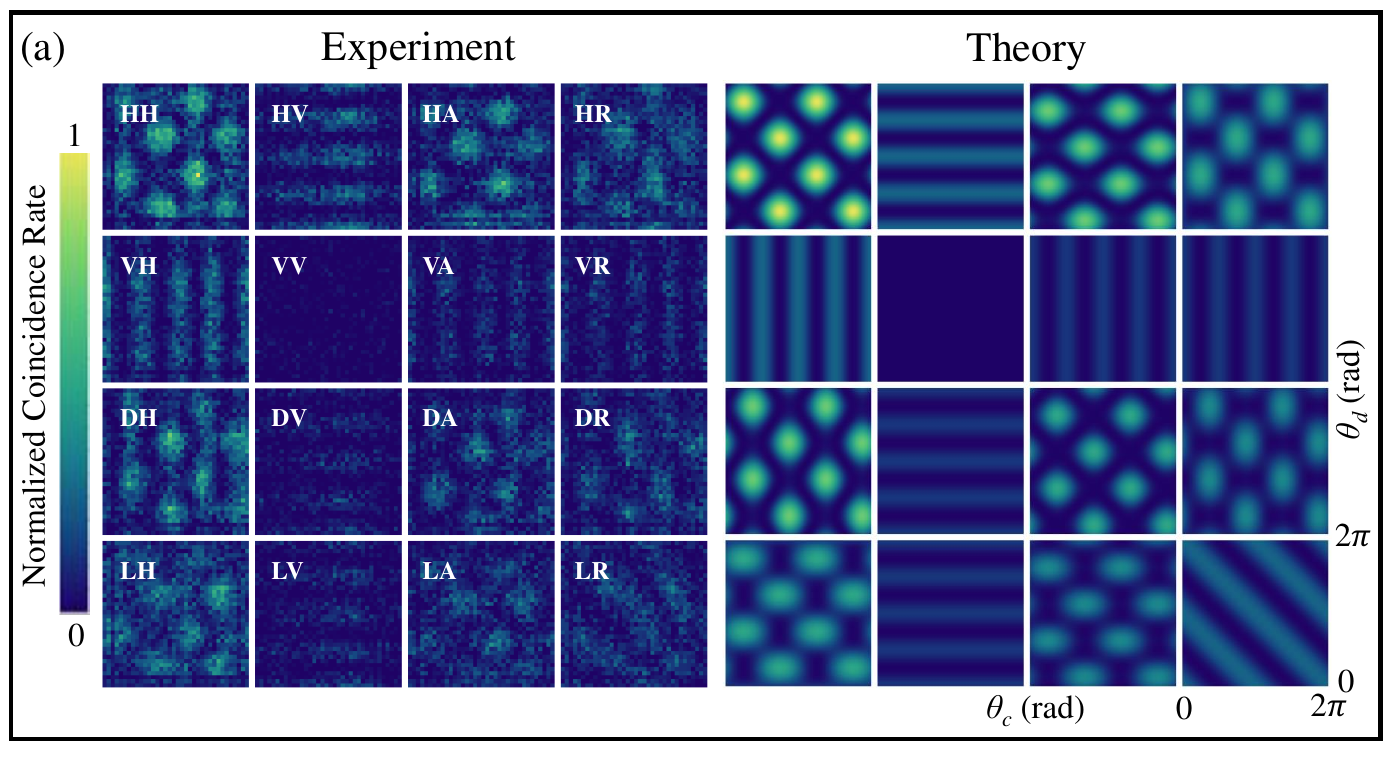}
\includegraphics [width= 1\textwidth]{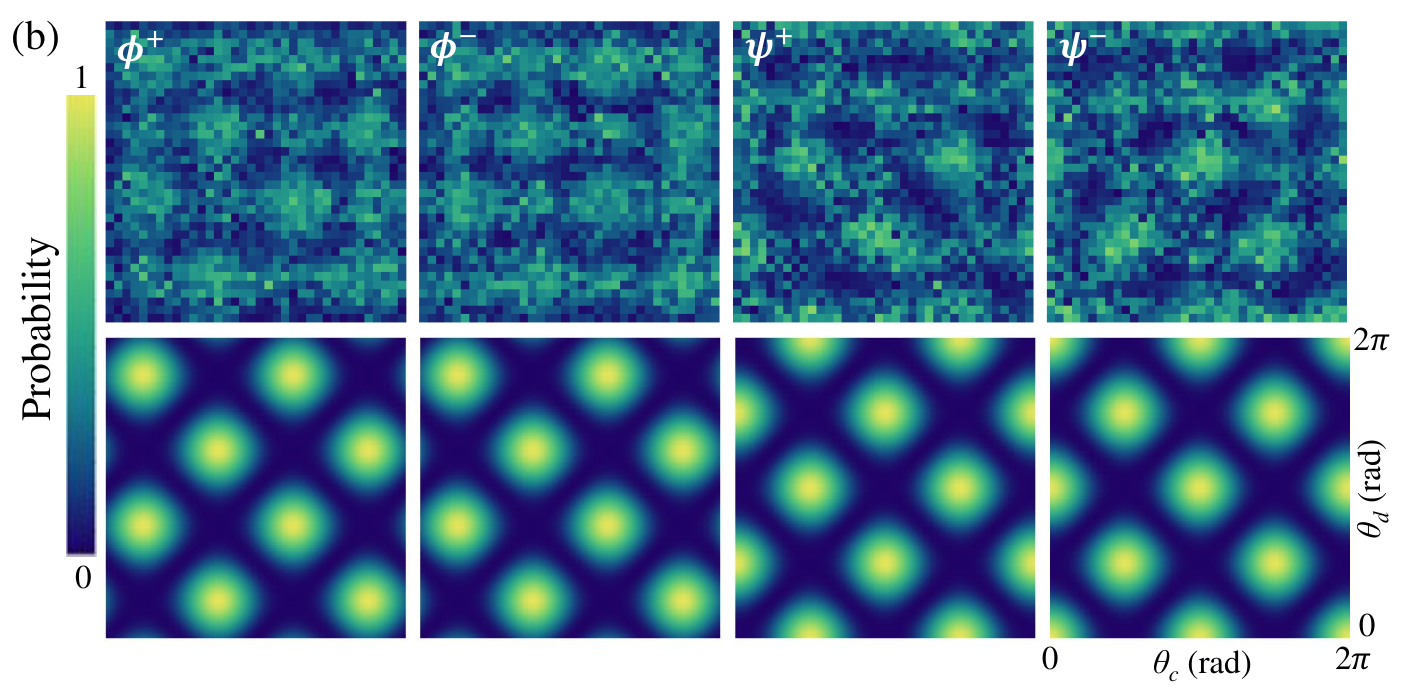}
\centering
\caption{\textbf{Generation of Bell states from HOM interference of VMs.} (a)~Experimental (left) and simulated (right) azimuthal correlations of the two-photon state for the 16 projections required for the complete state tomography. The topological charge of the first $q$-plate is ${q_a = 1}$ while the second $q$-plate is turned off. The initial state is $\hat{a}_{V}^\dagger\hat{b}_{H}^\dagger\ket{0}$. (b)~Experimental (top) and simulated (bottom) spatial patterns of the four Bell states ${\lbrace \ket{\phi^+},\ket{\phi^-},\ket{\psi^+},\ket{\psi^-} \rbrace}$ obtained via HOM interference of VMs in the case study. 
}
\label{case4}
\end{figure*}

\begin{figure*}[tbph]
\includegraphics [width= 1\textwidth]{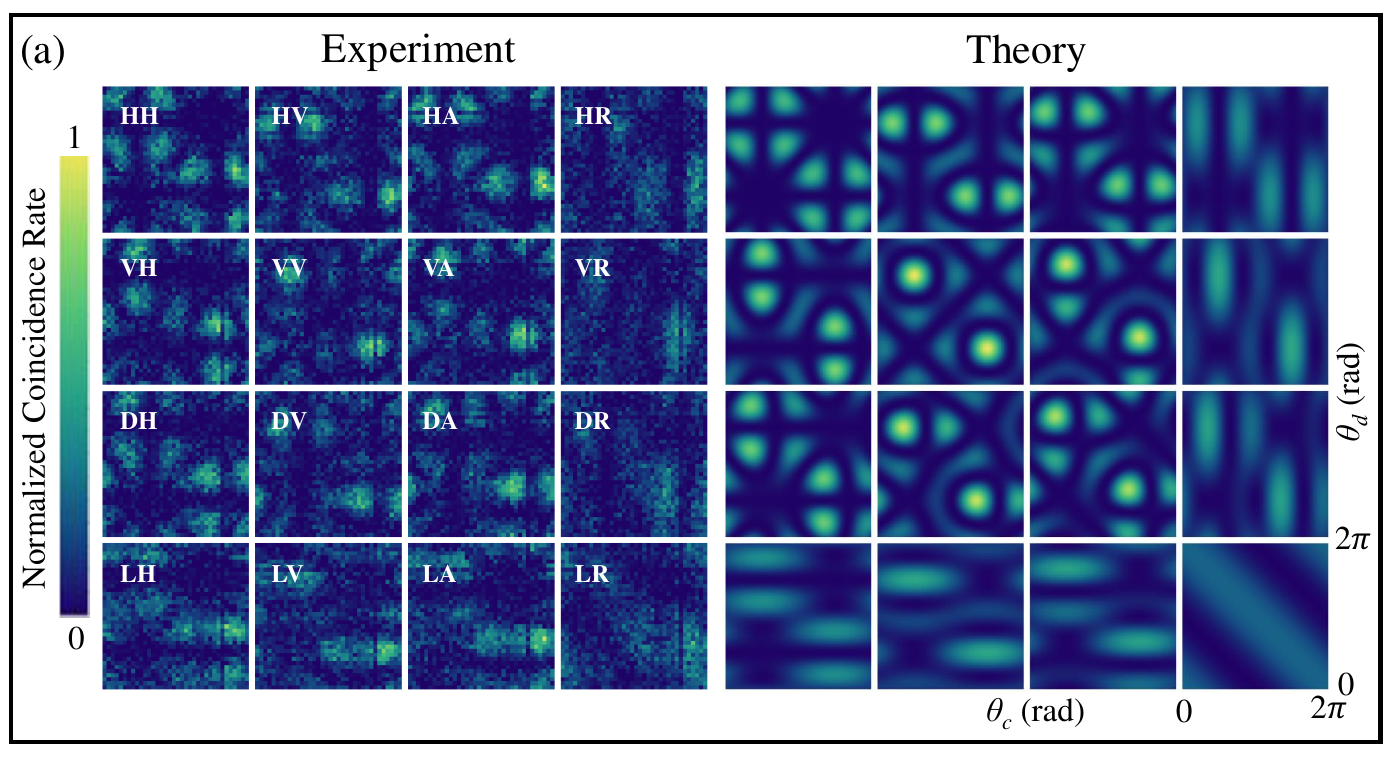}
\includegraphics [width= 1\textwidth]{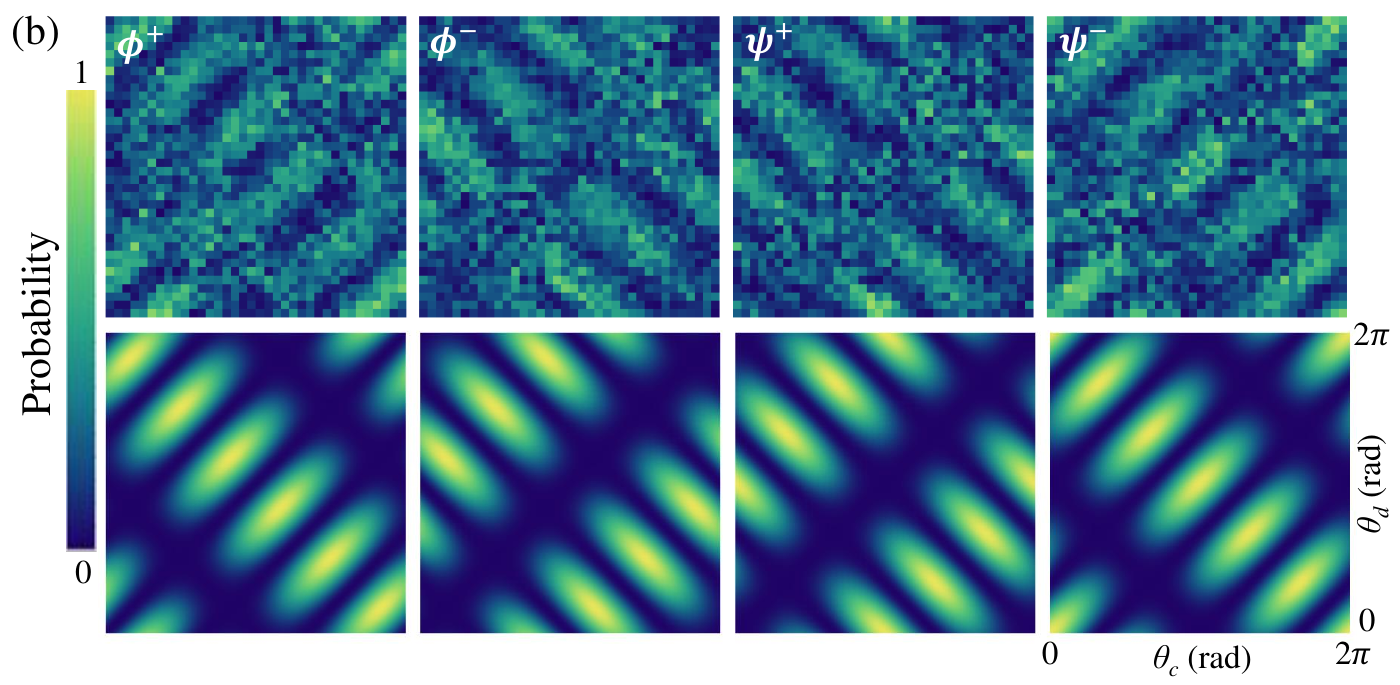}
\centering
\caption{\textbf{Generation of Bell states from HOM interference of VMs.} (a)~Experimental (left) and simulated (right) azimuthal correlations of the two-photon state for the 16 projections required for the complete state tomography. The topological charges of the two $q$-plates are ${q_a = 1}$ and ${q_b = 1/2}$. The initial state is $\hat{a}_{V}^\dagger\hat{b}_{H}^\dagger\ket{0}$. (b)~Experimental (top) and simulated (bottom) spatial patterns of the four Bell states ${\lbrace \ket{\phi^+},\ket{\phi^-},\ket{\psi^+},\ket{\psi^-} \rbrace}$ obtained via HOM interference of VMs in the case study. 
}
\label{case5}
\end{figure*}

In Figs.~\ref{case1}-\ref{case2}-\ref{case3}-\ref{case4}-\ref{case5}, we provide additional experimental results for different input VMs, obtained from various input polarizations and $q$-plates with different settings. 

%

%

%

%
%
%

\end{document}